\documentstyle[prl,aps,multicol,epsf,epsfig]{revtex}

\newcommand{\be}{\begin{equation}}
\newcommand{\ee}{\end{equation}}
\newcommand{\ba}{\begin{eqnarray}}
\newcommand{\ea}{\end{eqnarray}}
\newcommand{\rprl}[3]{Phys. Rev. Lett. {\bf #1}, #2 (19#3)}

\begin{document}

\title{\hfill {\small ULDF-TH-1/4/99} \\ \hfill \phantom{.} \\ 
Maximal Stability Regions for Superconducting 
Ground States\\ of Generalized Hubbard Models}  

\author{Matteo Beccaria}  

\address{ matteo.beccaria@le.infn.it \\ 
Dipartimento di Fisica, Universit\`a di Lecce, 
Via Arnesano, 73100 Lecce, Italy }  

\date{submitted to Physical Review Letters, April 9, 1999}  

\maketitle  

\begin{abstract} 
For a class of generalized Hubbard models, we determine the 
maximal stability region for the superconducting $\eta$-pairing 
ground state. We exploit the Optimized Ground State (OGS) approach 
and the Lanczos diagonalization procedure to derive a sequence of 
improved bounds. We show that some pieces of the stability boundary 
are asymptotic, namely independent on the OGS cluster size. In this 
way, necessary and sufficient conditions are obtained to realize 
superconductivity in terms of an $\eta$-pairing ground state. 
The phenomenon is explained by studying the properties of certain 
exact eigenstates of the OGS hamiltonians.
\end{abstract}


\begin{multicols}{2}
\narrowtext


Generalized Hubbard models are important theoretical frameworks for
the study of superconductivity. Apart from special cases, they are not
solvable and rigorous results on their physical properties are quite
valuable.

The inclusion of nearest neighbour extra couplings beyond the on site
Coulomb interaction has a long history and is still an interesting
issue.  Indeed, the qualitative effects of these interactions is not
definitely understood and examples can be provided where they are not
negligible at all.

As is well known, a good marker for superconductivity is off-diagonal
long-range order (ODLRO)~\cite{ODLRO}, a property that makes sense in
any number of dimensions and implies both Meissner effect and flux
quantization.  Ground state ODLRO can be detected by studying the
asymptotic behaviour of fermion correlation functions~\cite{NPODLRO}.
Of course, if the ground state is analitically known, it can be
checked explicitely.  This is the case of the so-called
$\eta$-pairing~\cite{Yang89} states that exhibit ODLRO and, under some
constraints, can be the ground states of certain generalized Hubbard
models.

When an $\eta$-pairing state is discovered to be an exact eigenstate,
the next problem is to determine the region in the coupling space
where it is also the ground state.  To answer this question many
analytical methods have been developed to establish rigorous bounds
for the superconducting region. Among them, we recall the positive
semidefinite operator approach~\cite{Positive,Arianna} and the bounds
derived by application of Gerschgorin's
theorem~\cite{Gerschgorin}. The algorithm which however appears to be
the simplest and most powerful is the Optimal Ground State (OGS)
scheme proposed for generalized Hubbard models~\cite{Boer95} and
recently applied to the case of next to nearest neighbour
couplings~\cite{Szabo98}.  The method is based on the exact
diagonalization of a certain local hamiltonian defined over a cluster
of sites. If the cluster is made larger, the superconducting region is
generally expected to expand. In the limit of an infinite cluster we
obtain exact bounds.

For simplicity, in the following we shall call superconducting (SC)
region, the subset of coupling space where the $\eta$-pairing state is
the ground state.  In this Letter, we apply the OGS algorithm to study
the stability of the superconducting $\eta$-pairing state with
momentum $\pi$.  We discuss the behaviour of the OGS bounds as a
function of the cluster size using the Lanczos algorithm to
diagonalize the cluster hamiltonian.  We obtain an improved SC region
that can be considered {\it numerically} asymptotic and discuss in
details the inclusion problem by stating the conditions under which
larger clusters are expected to give better bounds.  Another
interesting result is that some pieces of the boundary between the SC
and non SC regions are independent on the cluster size and determine
necessary and sufficient conditions for $\eta$-pairing
superconductivity.  We explain these stable boundaries by means of
certain exact eigenstates of the OGS hamiltonians whose properties are
crucial in this respect.

Let us consider the Hamiltonian of a one dimensional generalized
Hubbard model (we denote by $\langle i,j\rangle$ the sum over nearest
neighbour sites) \ba
\label{ham}
\lefteqn{H = -t\sum_{\langle i, j\rangle, \sigma=\uparrow, \downarrow}
(c^\dagger_{i\sigma} c_{j\sigma} + c^\dagger_{j\sigma} c_{i\sigma}) +}
&& \nonumber \\ &+& X\sum_{\langle i, j\rangle , \sigma =\uparrow,
\downarrow} (n_{i,-\sigma}+n_{j, -\sigma}) (c^\dagger_{i\sigma}
c_{j\sigma} + c^\dagger_{j\sigma} c_{i\sigma}) + \nonumber \\ &+&
U\sum_i \left(n_{i\uparrow}-\frac 1
2\right)\left(n_{i\downarrow}-\frac 1 2\right) + \nonumber \\ &+&
V\sum_{\langle i, j\rangle} (n_i-1)(n_j-1) + Y\sum_{\langle i, j,
\rangle} (p_i^\dagger p_j + p_j^\dagger p_i) +\nonumber \\ &+&
\frac{1}{2} J_{xy} (S^+_i S^-_j + S^+_j S^-_i) + J_z S^Z_i S^Z_j , \ea
where $c_{i\sigma}$ and $c^\dagger_{i\sigma}$ are canonical Fermi
operators obeying $\{c^\dagger_{i\sigma}, c_{j\sigma'}\} =
\delta_{ij}\delta_{\sigma\sigma'}$ and $\{c_{i\sigma},
c_{j\sigma'}\}=\{c_{i\sigma}^\dagger, c_{j\sigma'}^\dagger\}=0$. The
number operators are defined as usual:
$n_{i\sigma}=c^\dagger_{i\sigma} c_{i\sigma}$ and
$n_i=n_{i\downarrow}+n_{i\uparrow}$. The operator ${p_i}^\dagger$
creates pairs $p_i^\dagger = c_{i\uparrow}^\dagger
c_{i\downarrow}^\dagger$.  The Heisenberg exchange interaction is
written as usual in terms of $su(2)$ operators $S_i^+ =
c^\dagger_{i\uparrow} c_{i\downarrow}$, $S^- = (S^+)^\dagger$ and $S^Z
= \frac{1}{2} (n_\uparrow-n_\downarrow)$.

The Hamiltonian in~(\ref{ham}) contains many couplings: $X$
parametrizes the bond-charge repulsion interaction which has been
related to high-$T_c$ materials~\cite{BondCharge}; $U$ is the usual on
site Coulomb interaction; $V$ is the nearest neighbour charge-charge
coupling; $Y$ controls the pair hopping term as in the
Penson-Kolb-Hubbard models~\cite{PKH}.  Finally, $J_{xy}$ and $J_z$
are the $t-J$ like Heisenberg exchange couplings.

We introduce the $\eta$-pairing operator \be \eta^\dagger = \sum_n
(-1)^n p_n^\dagger , \ee from which we build the state \be
|\eta\rangle = (\eta^\dagger)^{N/2} |0\rangle , \ee where $|0\rangle$
is the empty state and $N$ is the number of lattice sites.  The state
$|\eta\rangle$ is an eigenstate of $H$ provided $V+2Y = 0$. In this
case it describes a half-filled state with energy \be E_+ =
\frac{1}{4}(U+4V) , \ee and can be shown to possess ODLRO.  Since
$E_+$ is an upper bound for the ground state energy, a strategy to
proof that $|\eta\rangle$ is the ground state is to find a lower bound
$E_-$ and a region in the coupling space where $E_-=E_+$.  The same
procedure applies also to other exact eigenstates like, for instance,
the $\eta$-pairing state with zero momentum.  Also, if
$V=Y=J_{xy}=J_z=0$ and $t=X$, exact eigenstates can be built by
applying $\eta^\dagger$ to eigenstates of the $U=+\infty$ standard
Hubbard model due to the fact that $t=X$ implies a conserved number of
doubly occupied sites.  However, we remark that the state
$|\eta\rangle$ is particularly interesting from a methodological point
of view as can be seen by the complete failure of the naive
Gerschgorin approach~\cite{Fail}.

Lower bounds for the ground state of $H$ may be obtained following the
OGS approach~\cite{Boer95}. The Hamiltonian~(\ref{ham}) is written as
\be
\label{split}
H = \sum_{n=-\infty}^\infty (h^{(1)}_n + h^{(2)}_{n, n+1}) , \ee where
$h_n^{(1)}$ contains operators acting only on site $n$ and $h_{n,
n+1}^{(2)}$ links site $n$ to site $n+1$ and depends on operators
acting on both.  To recast~(\ref{split}), we introduce extended
operators \ba {\tilde h}^{(k)}_n &=& \frac 1 2 h_n^{(1)} +
\sum_{m=n}^{n+k-2} h^{(2)}_{m, m+1} + \nonumber \\ &+&
\sum_{m=n+1}^{n+k-2} h_m^{(1)} + \frac 1 2 h^{(1)}_{n+k-1} , \ea for
any integer $k\ge 2$. The local hamiltonian $\tilde h^{(k)}$ describes
a cluster of $k$ sites.  Like $H$, also $\tilde h^{(k)}$ is symmetric
under $\eta^\dagger$. All the states \be (\eta^\dagger)^p
|\underbrace{0\cdots 0}_{\mbox{k sites}}\rangle , \quad p\
\mbox{integer} \ee are degenerate with energy $E_+$ and are precisely
those needed to build the $|\eta\rangle$ state on the infinite lattice
(see~\cite{Boer95} for a complete discussion of the $k=2$ case).  The
Hamiltonian can be written in terms of $\tilde h^{(k)}$ as \be
\label{decomp}
H = \sum_{n=-\infty}^{\infty} {\tilde h}^{(k)}_n .  \ee If we denote
by $E_0(N)$ the ground state energy for a system of $N$ sites, the
asymptotic ground energy per site is by definition \be {\cal E}_0 =
\lim_{N\to \infty} \frac{E_0(N)}{N} , \ee and, for each $k$, satisfies
the rigorous bound \be
\label{rbound}
{\cal E}_0 \ge \frac{1}{k-1} \min\sigma({\tilde h}^{(k)})
\stackrel{def}{=} {\cal E}^{(k)}_0 , \ee where $\sigma(A)$ denotes the
spectrum of the operator $A$.  The normalization factor $1/(k-1)$
takes into account the number of terms in~Eq.~(\ref{decomp}) which
contain a given site.  The right hand side of~Eq.~(\ref{rbound})
depends on the cluster size $k$ and a better bound is expected as $k$
increases. However, strictly speaking, this is false. Let us write a
cluster of $k$ sites in terms of smaller clusters \be
\label{dec}
\tilde h^{(k)}_n = \tilde h_n^{(k-l)} + \tilde h_{n+k-l-1}^{(l+1)},
\qquad 1\le l\le k-2 .  \ee From~(\ref{dec}) we obtain the exact
inequalities \be {\cal E}_0^{(k)} \ge \frac{k-l-1}{k-1} {\cal
E}_0^{(k-l)} + \frac{l}{k-1} {\cal E}_0^{(l+1)} , \ee and in
particular \be
\label{sequence}
{\cal E}_0^{(2k-1)} \ge {\cal E}_0^{(k)} .  \ee This allows to build
sequences of bounds converging to the exact bound in the infinite
cluster size limit.  In more details, Eq.~(\ref{sequence}) splits into
disjoint sequences of cluster sizes as follows \ba
\label{tree}
2\subset 3\subset 5\subset\cdots && ,\nonumber \\ 4\subset 7\subset
13\subset\cdots && , \ea where the notation means that each sequence
gives better and better bounds. We notice that in general two
sequences are not related at all and, in particular, $3\subset 4$ may
be false as we shall see in explicit examples.  What can be stated in
full generality is that the minimal choice $k=2$ is always the worst
bound since from \be {\cal E}_0^{(N+1)} \ge \frac{N-1}{N} {\cal
E}_0^{(N-1)} + \frac{1}{N} {\cal E}_0^{(2)} , \ee we proof inductively
that $\forall N\ge 2$ we have \be {\cal E}_0^{(N)} \ge {\cal
E}_0^{(2)} .  \ee Keeping these remarks in mind, we study the size
dependence of the conditions under which~(\ref{ham}) with $Y=-2V$
admits $|\eta\rangle$ as its ground state by explicit diagonalization
of $\tilde h^{(L)}$ on clusters of increasing sizes. The OGS method
requires diagonalization of the local hamiltonian in all sectors of
definite up and down electron numbers; for the numerical
diagonalization we use the Lanczos algorithm.  In the following we
shall always assume~$t\equiv 1$ and denote by $L$ the cluster size.

Since the couplings constants space is large, we decide to discuss
separately what happens with the Heisenberg exchange interaction
switched on or off. Let us begin with $J_{xy} = J_z = 0$.

In Fig.~1 we show at $X=0$ the size dependence of the bounds when
$V>-1$. As can be seen, there are regions where the corrections are
definitely negligible beyond $L=3$, i.e. $V>-0.4$. On the other hand,
around $V=-0.5$, size effects can be important up to large cluster
sizes. We remark that this figure does not show any non trivial
relationship among the bounds obtained at different $L$: they just
improve monotonically.

In Fig.~2 we show the best bounds obtained with $L=6$ at several
values of $X$. In the inset we expand the region around $V=-2$. A
remarkable feature of the plot is that an enveloping straight line
appears around $V=-1$. Successive corrections are quite small and the
region shown can be considered maximal from any practical point of
view~\cite{Data}.

In Fig.~3 we plot at four different $X$ the difference $\Delta
U(L)=U(L)-U(2)$ between the boundary curves at $L>2$ and the minimal
one at $L=2$.  The inclusion tree~(\ref{tree}) is non trivially
satisfied and indeed the $L=4$ bound is not always better than the
$L=3$ one.  As a second remark, we observe that at each $X$ there is a
piece of the boundary where finite size corrections vanish. This turns
out to happen between two of the $L=2$ boundary points. As shown
in~\cite{Boer95}, the result at $L=2$ is that a sufficient condition
for $|\eta\rangle$ to be the the ground state is \ba V &\le& 0 , \\ U
&\le& -2\ \mbox{max}\left(2+2V, 2|1-2X|+2V, V-\frac{(1-X)^2}{V}\right)
.  \nonumber \ea For $0\le X\le 1$ (we study this case only), the
difference $\Delta U$ vanishes between the intersections of the curves
$U=-4(1+V)$ and $U=-2(V-(t-X)^2/V)$, namely for $|V+1|\le
\sqrt{X(2-X)}$.

Let us now discuss why this stable boundary subset appears. For each
value of $L$, the normalized cluster Hamiltonian $\frac{1}{L-1} \tilde
h^{(L)}$ has many eigenstates $|E^{(L)}_i(U, V, X)\rangle$
($i=1,\dots, \dim(\tilde h)$) which we label by their eigenvalue.

Let $X$ play the role of a parameter; following the OGS approach, the
inequalities $E^{(L)}_i\ge E_+$ determine the superconducting region
in the $(U, V)$ plane.  Each point of its boundary satisfies
$E_i^{(L)}=E_+$ for some index $i$. Hence, if a subset of the boundary
turns out to be $L$ independent, a possible reason can be the
existence of an eigenvalue independent on $L$. A trivial case is
provided by the states $(\eta^\dagger)^p|0\rangle$ ($p$ integer) where
$|0\rangle$ is the empty state for $\tilde h^{(L)}$. However, in this
case, the condition $E^{(L)}=E_+$ is identically satisfied for all
$U$, $V$ and $X$ and does not determine any boundary.  To find a non
trivial eigenstate with eigenvalue independent on $L$ we can consider
the one particle sector (i.e. $n_\uparrow = 1$, $n_\downarrow = 0$ or
viceversa). The two states \be |S_\sigma\rangle = \sum_{n=1}^L
c_{n\sigma}^\dagger |0\rangle,\qquad \sigma = \uparrow, \downarrow ,
\ee are indeed exact eigenstates of $\frac{1}{L-1} \tilde h^{(L)}$
provided $U = -4(1+V)$ and in this case their eigenvalue is precisely
$E_+=-1$ ($t\equiv 1$).  The states $|S_\sigma\rangle$ are thus
responsible for the stable boundary. To understand why it is confined
to $|V+1| \le \sqrt{2X-X^2}$ we introduce additional eigenstates of
$\tilde h^{(L)}$.  Indeed, on the line $U=-4(V+1)$, the $su(2)$
singlet state ($X\neq 1$) \be
\label{gamma}
|\gamma\rangle = \left\{ \sum_{i\neq j} c^\dagger_{i\uparrow}
c^\dagger_{j\downarrow} + \rho \sum_i \frac{1-(-1)^{i+L}}{2}
p_i^\dagger \right\} |0\rangle , \ee can be shown to be an exact
eigenstates of $\frac{1}{L-1}\tilde h^{(L)}$ with eigenvalue $E_+$ if
and only if $\rho = (2+V)/(1-X)$ and $V=-1\pm\sqrt{2X-X^2}$. This is
the $L>2$ generalization of the state $|\psi_\pm\rangle$ discussed
in~\cite{Boer95} in the $L=2$ case.  It forbids to extend the bounds
associated to $|S_\sigma\rangle$ beyond the points
$|V+1|=\sqrt{2X-X^2}$. The case $X=1$ is singular and must be treated
separately; the number of doubly occupied sites is conserved and
splitting may occur. For instance, the state $|\gamma_+\rangle$ with
$V=0$ splits into the independent eigenstates $|\gamma_i\rangle =
p_i^\dagger |0\rangle$.

The above analytical and numerical results lead us to the conclusion
that the inequality $U\le -4(1+V)$ is a necessary and sufficient
condition for $|\eta\rangle$ being the ground state in the subset of
the coupling space constrained by the conditions $t\equiv 1$,
$Y+2V=0$, $0<X<1$ and $|V+1| \le \sqrt{X(2-X)}$.

The above facts do not change qualitatively when the Heisenberg
exchange interaction is switched on. To simplify the analysis, we
consider the special point $J_{xy}=J_z=-2Y$ which allows for a
comparison with the results illustrated in~\cite{Arianna,Boer95}.

The plot of Fig.~4 is analogous to that of Fig.~1; we show the OGS
bounds in the $(U,V)$ plane at $X=0$. A part of the boundary is
clearly independent on $L$. This fact is clearly visible in the left
inset and mostly in Fig.~5 where (as in Fig.~3) the asymptotic part of
the boundary can be seen as a function of $X$. In particular, the left
edge is at $V=-1$ and is independent on $X$. These features can be
analyzed as in the previous case.  The bounds valid at $L=2$ are \ba V
&\le& 0 , \\ U &\le& -2\ \mbox{max}\left(0, 2+2V, 2|1-2X|+2V,
4V-\frac{(1-X)^2}{V}\right) , \nonumber \ea and the stable part (with
$0\le X\le 1$) is that between the intersections of the lines $U=0$,
$U=-4(1+V)$ and the curve $U = -2(4V-(1-X)^2/V)$. Thus we obtain the
following interval for $V$: $-1\le V\le \frac{1}{2}
(1-\sqrt{3-4X+2X^2})$.  As before, the line $U=-4(1+V)$ appears at all
$L$ since on it the states $|S_\sigma\rangle$ are eigenstates of
$\frac{1}{L-1}\tilde h^{(L)}$. Moreover, as before, there are two
states which forbid to cross the above interval for $V$.  At the right
edge, such a state has the form of~Eq.~(\ref{gamma}) with $\rho =
2(1-V)/(1-X)$ and $V=\frac 1 2 (1-\sqrt{3-4X+2X^2})$. At the left
edge, $V=-1$, the exact ($X$ independent) eigenstate is instead \be
|\gamma'\rangle = \left( \sum_{i<j} c^\dagger_{i\uparrow}
c^\dagger_{j\downarrow} +\sum_{i>j} c^\dagger_{j\downarrow}
c^\dagger_{i\uparrow} \right) |0\rangle .  \ee To summarize, in this
case, our conclusion is that under the conditions $t=1$, $Y+2V=0$,
$0<X<1$ and $-1\le V \le \frac{1}{2} (1-\sqrt{3-4X+2X^2})$, a
necessary and sufficient condition for $|\eta\rangle$ being the ground
state is $U\le -4(1+V)$.

To conclude, in this Letter we have considered generalized Hubbard
models with nearest neighbour couplings and the problem of determining
when the ground state is a superconducting $\eta$-pairing state. By
diagonalizing local hamiltonians associated to clusters of sites with
different sizes we studied the convergence of the OGS bounds. As
predicted in~\cite{Boer95}, it may happen that the bounds obtained
with the smallest clusters are actually exact.  This peculiar
situation seems rather typical and indeed we have shown that there
exist subsets of the bounding region which are asymptotic and remain
unchanged as the cluster size is varied.  We clarified the origin of
the phenomenon by providing several exact eigenstates which play a
crucial role in its derivation.

We thank A. Montorsi for very useful discussions on generalized
Hubbard models and $\eta$-pairing superconductivity.

\end{multicols}

\widetext

\newpage

\begin{figure}[htb]
\centerline{\hbox{\epsfig{figure=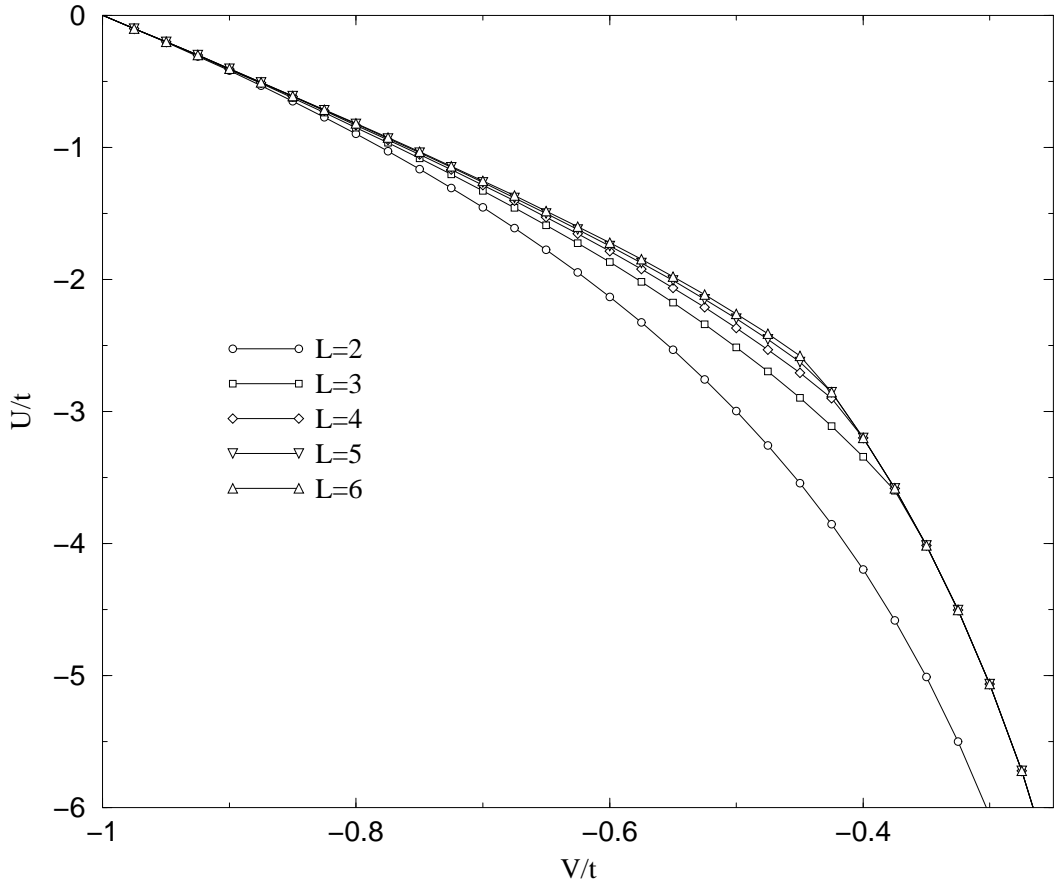,width=12cm,angle=-90}}}
\vskip 0.5cm
\caption{Size dependence of the OGS bounds in the $(U,V)$ plane at
$J_{xy}=J_z=0$ and $X=0$.}
\end{figure}\noindent

\newpage

\begin{figure}[htb]
\centerline{\hbox{\epsfig{figure=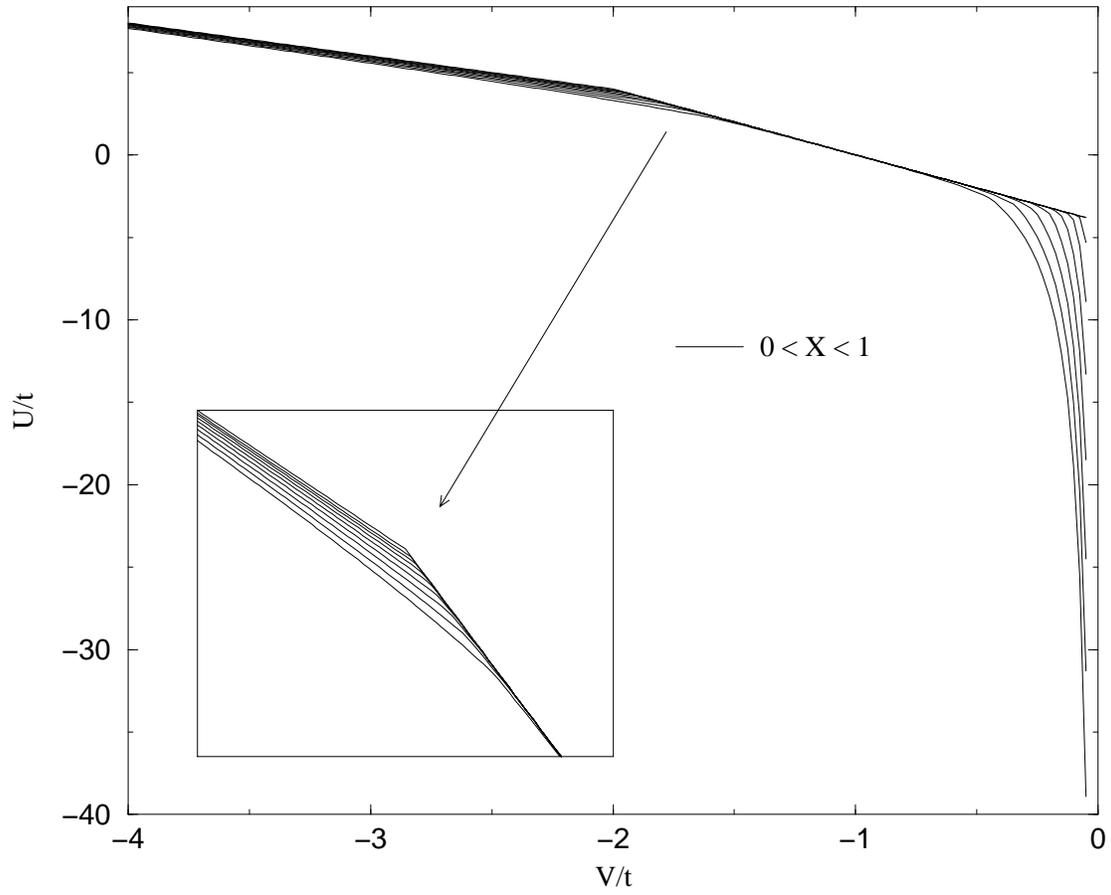,width=12cm,angle=-90}}}
\vskip 0.5cm
\caption{Best OGS bounds obtained with $L=6$ clusters. The Heisenberg
interactions are switched off $J_{xy}=J_z=0$. The different curves
correspond (from bottom to top) to $X=0$, $0.1$, $0.2$, $0.3$, $0.4$,
$0.5$, $0.6$, $0.7$, $1.0$.  }
\end{figure}\noindent

\begin{figure}[htb]
\centerline{\hbox{\epsfig{figure=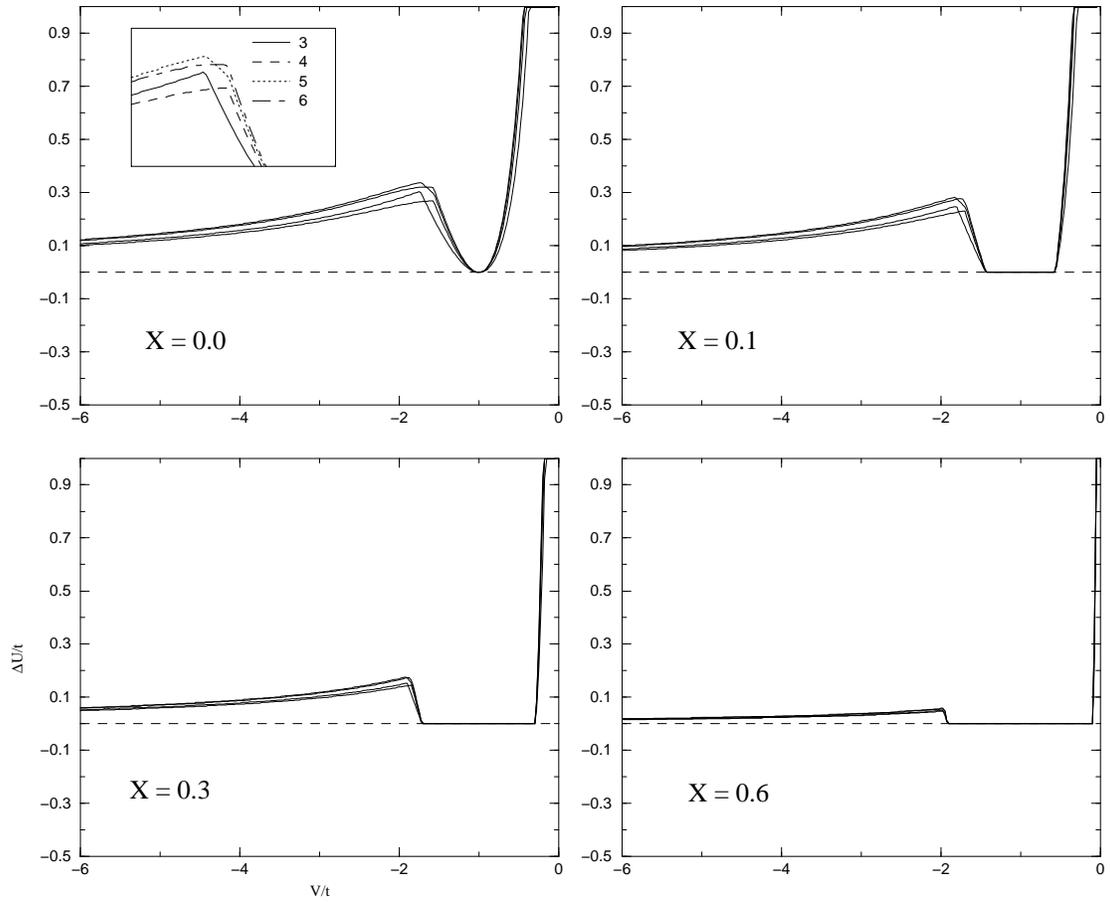,width=12cm,angle=-90}}}
\vskip 0.5cm
\caption{Existence of an asymptotic boundary. The plots show $U(V;
L)-U(V; 2)$ as a function of $V$ (always in units of $t$) for four
values of $X$. The function $U(V; L)$ is the curve obtained from the
OGS bounds using clusters of $L$ sites.  The inset at $X=0$ shows a
non trivial inclusion tree as the cluster size is increased.  See also
Fig.~4.  }
\end{figure}\noindent

\begin{figure}[htb]
\centerline{\hbox{\epsfig{figure=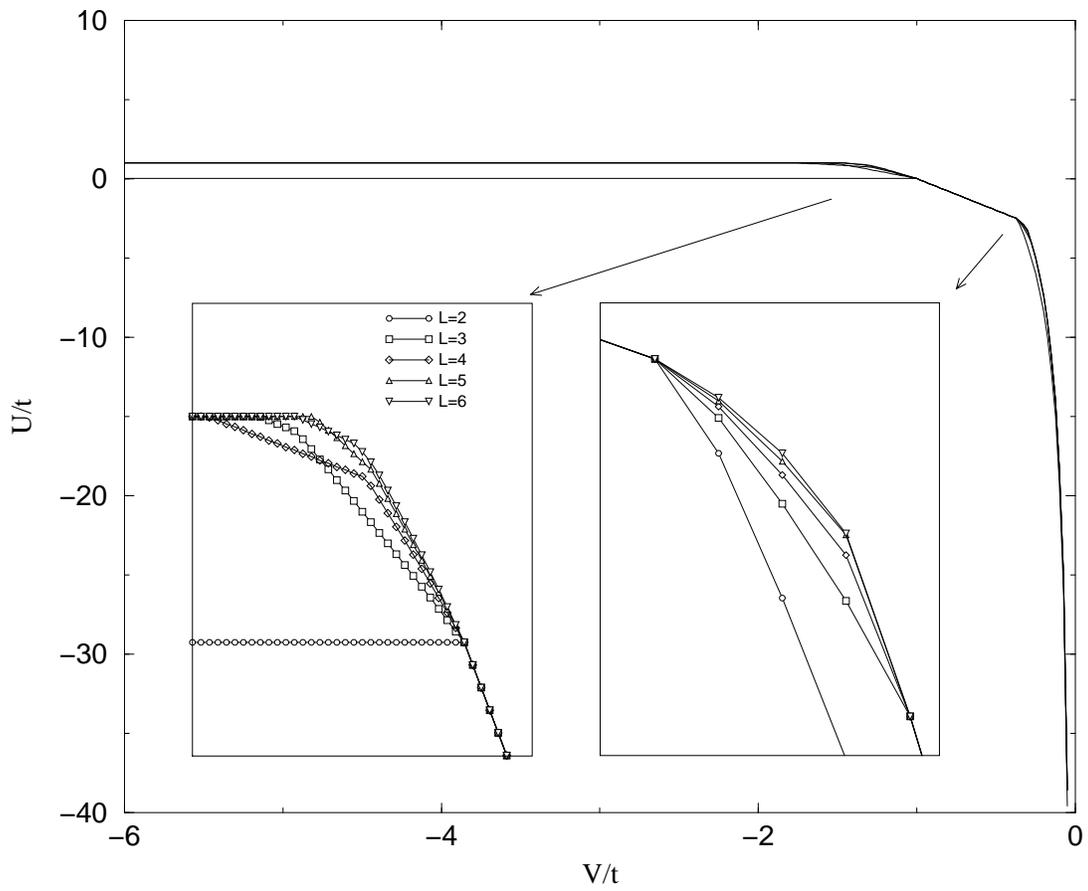,width=12cm,angle=-90}}}
\vskip 0.5cm
\caption{As Fig.~1, but with $J_{xy}=J_z=-2Y$. The left inset shows a
non trivial inclusion tree.}
\end{figure}\noindent

\begin{figure}[htb]
\centerline{\hbox{\epsfig{figure=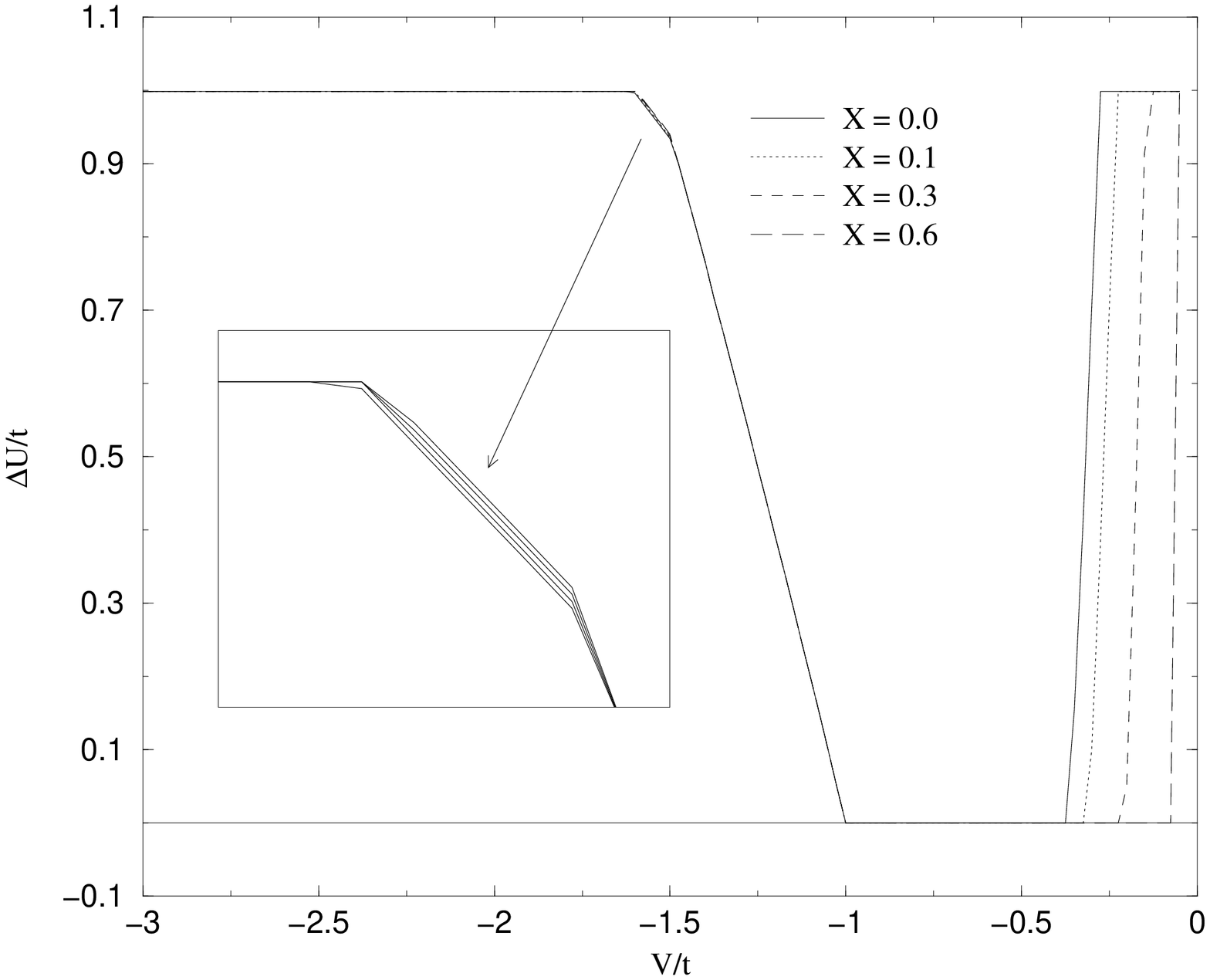,width=12cm,angle=-90}}}
\vskip 0.5cm
\caption{As Fig.~4, but with $J_{xy}=J_z=-2Y$. The inset magnifies a
portion of the curves to confirm that the OGS bound is not asymptotic
for $V < -1$.  }
\end{figure}\noindent


\begin{references}




\bibitem{ODLRO} C. N. Yang, Rev. Mod. Phys. {\bf 34}, 694 (1962);
G. L. Sewell, J. Stat. Phys. {\bf 61}, 415 (1990).


\bibitem{NPODLRO} R. Friedberg and T. D. Lee, Phys. Rev. B {\bf 40},
6745 (1989); H. Q. Lin, E. R. Gagliano, D. K. Campbell, E. H. Fradkin
and J. E. Gubernatis, in {\it The Hubbard Model}, Edited by
D. Baeriswyl et al., Plenum Press, New York (1995); G. Bouzerar and
G. I. Japaridze, Report No. cond-mat/9605161; R. T. Clay,
A. W. Sandvik, D. K. Campbell, Phys. Rev. B{\bf 59}, 4665 (1999)


\bibitem{Yang89} C. N. Yang, \rprl{63}{2144}{89}.


\bibitem{Putikka92} W. O. Putikka, M. U. Luchini and M. Ogata,
\rprl{69}{2288}{92}.


\bibitem{Positive} U. Brandt and A. Giesekus, \rprl{68}{2648}{92};
R. Strack and D. Vollhardt, \rprl{70}{2637}{93}; R. Strack and
D. Vollhardt, \rprl{72}{3425}{94}; R. Strack and D. Vollhardt, in {\it
The Hubbard Model}, Edited by D. Baeriswyl et al., Plenum Press, New
York (1995);


\bibitem{Arianna} A. Montorsi and D. K. Campbell, Phys. Rev. B{\bf
53}, 5153 (1996).


\bibitem{Gerschgorin} A. A. Ovchinnikov, Mod. Phys. Lett. B{\bf 7},
1397 (1993); A. A. Ovchinnikov, J. Phys. C{\bf 6}, 11057 (1994); J. de
Boer, V. E. Korepin and A. Schadschneider, \rprl{74}{789}{95}.


\bibitem{Boer95} J. de Boer and A. Schadschneider,
\rprl{75}{4298}{95}.

\bibitem{Szabo98} Z. Szab\'o, Report No. cond-mat/9807083. We point
out that due to a different normalization of the Heisenberg couplings,
the values of $J_{xy}$ and $J_z$ are doubled with respect to ours and
those of~\cite{Arianna,Boer95}.

\bibitem{BondCharge} J. E. Hirsch, Physica C{\bf 158}, 326 (1990);
R. Z. Bariev, A. Kl\"umper, A. Schadschneider, J. Zittartz, J. Physics
A{\bf 26}, 1249 (1993).

\bibitem{PKH} K. A. Penson, M. Kolb, Phys. Rev. B{\bf 33}, 1663
(1986); K. A. Penson, M. Kolb, J. Stat. Phys. {\bf 44}, 129 (1986);
I. Affleck, J. B. Marston, J. Physics C{\bf 21}, 2511 (1988).

\bibitem{Fail} The reason for the failure of the Gerschgorin approach
in the case of the $|\eta\rangle$ state is that for the state
$s=|\uparrow\downarrow, 0, \uparrow\downarrow, 0, \uparrow\downarrow,
0, \cdots\rangle$ the Gerschgorin lower bound is $E_- =
H_{ss}-\sum_{s'\neq s} H_{ss'} < E_+$ and we cannot never impose the
equality $E_-=E_+$

\bibitem{Data} The numerical data defining the boundary between the SC
and non SC region are available upon request to the author.

\end{references}
\end{document}